\documentclass[letterpaper,preprintnumbers,english,floats,floatfix,amssymb,prd,superscriptaddress,nofootinbib]{revtex4-2}
\usepackage{amssymb,amsmath}
\usepackage{epsfig,hyperref}
\usepackage[svgnames]{xcolor}
\usepackage{pgf,tikz}
\usepackage{epstopdf}
\usepackage[british]{babel}
\usepackage{enumerate}
\usepackage{enumitem}
\usepackage{makecell}
\usepackage{booktabs}
\usepackage{graphicx}
\usepackage{nicefrac}
\usepackage{units}
\usepackage{mathrsfs}
\usepackage{placeins} %

\makeatletter
\DeclareRobustCommand*{\bfseries}{%
	\not@math@alphabet\bfseries\mathbf
	\fontseries\bfdefault\selectfont
	\boldmath
}
\makeatother

\usepackage{graphicx}
\usepackage{dcolumn}
\usepackage{bm}
\usepackage{subfigure}
\usepackage{hyperref}
\usepackage{xcolor}
\begin{document}

\preprint{APS/123-QED}

\title{Stress-energy tensor of quantized scalar fields in thermal states on a zero-tidal wormhole}

\author{Shun Jiang}
\email{shunjiang@mail.bnu.edu.cn}
\author{Xiangdong Zhang}
\email{Corresponding author. scxdzhang@scut.edu.cn}
\affiliation{School of Physics and Optoelectronics, South China University of Technology, Guangzhou 510641, China\\}
\date{\today}

\begin{abstract}
The construction of a static traversable wormhole requires exotic matter that satisfies the Morris-Thorne conditions. Quantum energy-momentum tensors have long been considered the most promising candidate for such exotic matter. In this paper, we present the first calculation of the stress-energy tensor for a quantum massive scalar field in thermal states localized on the throat of a zero-tidal-force wormhole. By varying the dimensionless temperature and dimensionless mass of the scalar field, we find that the Morris-Thorne conditions can only be satisfied when the scalar field mass falls within a specific bounded interval. Furthermore, for any scalar field mass within this interval, there always exists a mass-dependent dimensionless critical temperature: the Morris-Thorne conditions are fulfilled only if the temperature remains below this critical threshold.

\end{abstract}


\maketitle

\section{INTRODUCTION}

The concept of hypothetical spacetime tunnels connecting two distinct asymptotic spacetime regions—universally known as wormholes—can be traced back to the pioneering early work of Einstein and Rosen \cite{Einstein:1935tc}. However, the systematic study of traversable wormholes, i.e., structures that can theoretically be crossed by macroscopic observers, was not formally established until the landmark work of Morris and Thorne \cite{Morris:1988cz}. They demonstrated that to sustain the wormhole geometry, matter localized at the wormhole throat must satisfy the Morris-Thorne conditions, which necessarily implies a violation of standard energy conditions. Given these exotic properties, the matter supporting the wormhole throat was termed exotic matter. Since quantum scalar fields are capable of violating energy conditions, Morris, Thorne, and Yurtsever subsequently proposed that such fields could be exploited to stabilize traversable wormholes. \cite{Morris:1988tu}.

Due to the requirement of renormalization \cite{Christensen:1976vb}, calculating the energy-momentum tensor of quantum scalar fields has long been a challenging problem. Using the Dewitt-Schwinger approximation, analytic approximate expressions for the stress-energy of a quantized scalar field are obtained in \cite{Anderson:1994hg,Anderson:1993if}. Using approximate expressions, whether quantum scalar fields can satisfy the Morris-Thorne conditions has been investigated in \cite{Taylor:1996yu,Popov:2005qy,Kocuper:2017aap,Popov:2014rya,Matyjasek:2020cmi,Popov:2001kk,Khusnutdinov:2003ii}. By substituting analytic approximate energy-momentum tensor expressions into the Einstein field equation, wormhole solutions supported by quantum scalar fields were obtained in \cite{Hochberg:1996ee}. To better investigate whether a quantum scalar field can support a wormhole, the exact quantum energy-momentum tensor under the short-throat approximation has been calculated in \cite{Bezerra:2010ix}. To overcome the difficulties encountered in the renormalization process, Levi and Ori proposed the pragmatic mode-sum regularization method \cite{Levi:2015eea,Levi:2016paz,Levi:2016esr}. This method has a wide range of applicability, it only requires spacetime to admit a Killing vector field, and has been used to calculate the quantum energy-momentum tensor for various black hole spacetimes \cite{Levi:2016quh,Levi:2016exv,Zilberman:2019buh,Zilberman:2022aum,Ori:2025zhe}. 

To date, existing studies of the quantum energy-momentum tensor in wormhole spacetimes have been restricted exclusively to the ground state. This gives rise to a fundamental open question: what is the actual quantum state of a dynamically formed wormhole? Insights from the spherical shell collapse model tell us that when a spherical shell collapses into a black hole from an initial vacuum state, the resulting quantum state is described by the Unruh state rather than the Boulware state \cite{Unruh:1976db}. This suggests that the quantum state of a wormhole may itself be determined by its formation history. Since no generally accepted framework for wormhole formation currently exists, the exact quantum state characterizing a macroscopic wormhole remains unresolved. However, for a static wormhole admitting a timelike Killing vector field, there are two important classes of quantum states that are invariant under the action of the Killing isometry: the ground state and the thermal equilibrium state. \cite{Sahlmann:2000fh}. 

To the best of our knowledge, no existing work has yet investigated whether the exact quantum energy-momentum tensor of a thermal quantum state can act as the exotic matter supporting a traversable wormhole. This motivates the core question addressed in this work: can a stable traversable wormhole exist in a thermal quantum state? In this paper, we present the first calculation of the renormalized energy-momentum tensor for a massive quantum scalar field in a thermal state, evaluated at the throat of a zero-tidal-force wormhole—the simplest class of traversable wormholes with vanishing radial tidal forces. We work within the Hadamard renormalization framework \cite{Decanini:2005eg}, and adopt the mode-sum regularization scheme developed by Levi and Ori \cite{Levi:2015eea,Levi:2016paz,Levi:2016esr} to compute the renormalized stress-energy tensor. By varying the temperature of the thermal state and the mass of the scalar field, we systematically explore whether there exists a parameter regime where the quantum energy-momentum tensor satisfies the Morris-Thorne conditions.

This article is divided as follows: In Section \ref{section2}, we give a brief review of the geometry of zero-tidal wormholes and the Morris-Thorne conditions. In Section \ref{section3}, we present the expression for the renormalized energy-momentum tensor of a quantum scalar field in a thermal state within the Hadamard renormalization framework. In Section \ref{section4}, we numerically calculate the renormalized quantum energy-momentum tensor for different temperatures and scalar field masses, and identify the parameter region that satisfies the Morris-Thorne conditions. In Section \ref{section5}, we present a detailed discussion of our results.

\section{ZERO-TIDAL WORMHOLES AND THE MORRIS-THORNE CONDITIONS}\label{section2}

In this section, we first review the general traversable wormhole metric presented in \cite{Morris:1988cz}. We then discuss the conditions that the energy-momentum tensor at the throat of a traversable wormhole must satisfy, namely the Morris-Thorne conditions. Finally, we introduce the zero-tidal wormhole model considered in this paper—the simplest traversable wormhole with zero radial tidal.

The line element of a general static, spherically symmetric wormhole can be expressed as follows

\begin{eqnarray}
ds^2=-e^{2\Phi(r)}dt^2+\left(1-\frac{b(r)}{r}\right)^{-1}dr^2+r^2d\Omega^2, \label{general metric form}
\end{eqnarray}
where $\Phi(r)$ and $b(r)$ are the redshift function and the shape function, respectively. For a traversable wormhole to exist, these functions must satisfy the following conditions

(S1) There exist an positive value $r_0$ such that $b(r_0)=r_0$, and the inequality $1-\frac{b(r)}{r}\geq0$  holds for $[r_0,\infty)$.

(S2) In the neighborhood of $r_0$, the inequality $\frac{d^2r}{dr_*^2}>0$ holds, where $r_*(r)=\pm\int_{r_0}^{r}\left(1-\frac{b(r)}{r}\right)^{-\frac{1}{2}}dr$ is the proper radial distance. Here, the positive and negative signs correspond to the space-time regions on either side of the wormhole throat respectively.

(S3) $\Phi(r)$ is globally finite throughout the spacetime.

Conditions (S1) and (S2) collectively imply the presence of a wormhole throat at the radial coordinate $r=r_0$. Condition (S3) guarantees the absence of any horizons. 

To satisfy these conditions, constraints must be imposed on the energy-momentum tensor at the wormhole throat, namely the Morris-Thorne conditions. For the sake of convenience, we first introduce a set of tetrads as follow
\begin{eqnarray}
	&&e^a_{0}=e^{-\Phi(r)}\left(\frac{\partial}{\partial t}\right)^a,\quad e^a_1=\left(1-\frac{b(r)}{r}\right)^{\frac{1}{2}}\left(\frac{\partial}{\partial r}\right)^a,\nonumber\\
	&&e^a_2=\frac{1}{r}\left(\frac{\partial}{\partial \theta}\right)^a,\quad e^a_3=\frac{1}{r\sin\theta}\left(\frac{\partial}{\partial \phi}\right)^a.
\end{eqnarray}
Substitute the metric (\ref{general metric form}) into Einstein's field equations $G_{ab}=8\pi T_{ab}$, and taking into account the conditions (S1)-(S3), the Morris-Thorne conditions can be expressed as follows
\begin{eqnarray}
	\tau_{0}:=-T_{11}(r_0)\ge 0, \label{MT1}\\
	\eta_0:=-T_{11}(r_0)-T_{00}(r_0)\ge 0 \label{MT2}.
\end{eqnarray}

From condition (\ref{MT1}), it can be observed that a positive value of $\tau_0$ indicates the presence of tension at the wormhole throat. Condition (\ref{MT2}) indicates that the energy-momentum tensor of the matter at the wormhole throat violates the energy conditions. Matter that satisfies condition (\ref{MT1}) and  condition (\ref{MT2}) is thus termed exotic matter.

In this paper, we will investigate whether the energy-momentum tensor of the thermal state of a quantum scalar field in a zero-tidal wormhole can serve as the exotic matter required to sustain the wormhole throat. ‌The line element of the zero-tidal wormhole‌ can be obtained by setting  $\Phi(r)=0$ and $b(r)=b_0$ in (\ref{general metric form}),  yielding

\begin{eqnarray}
ds^2=-dt^2+\left(1-\frac{b_0}{r}\right)^{-1}dr^2+r^2d\Omega^2.
\end{eqnarray}

\section{Quantum Field Theory in Wormhole SpaceTime}\label{section3}

In this section, we will present the expression of the renormalized energy-momentum tensor for the thermal state of a quantum massive scalar field.

The equation for a minimally coupled massive scalar field can be written as
\begin{eqnarray}
	\left(\nabla_a\nabla^a-m_0^2\right)\phi=0.
\end{eqnarray}
The classical stress-energy tensor is given by
\begin{eqnarray}
	T_{uv}=\phi_{;u}\phi_{;v}-\frac{1}{2}g_{uv}g^{\rho\sigma}\phi_{;\rho}\phi_{;\sigma}-\frac{1}{2}g_{uv}m_0^2\phi^2. \label{classical stress energy tensor}
\end{eqnarray}

Using spacetime symmetry, we can decompose the mode function in the form
\begin{eqnarray}
	\psi_{\omega lm}(t,r,\theta,\varphi)=\frac{1}{\sqrt{4\pi|\omega|}}e^{-i\omega t}Y_{l m}(\theta,\varphi)\frac{R_{\omega l}(r)}{r}. \label{mode decompose}
\end{eqnarray}
Here $Y_{l m}(\theta,\varphi)$ are the spherical harmonics functions and $R_{\omega l}(r)$ satisfies the radial equation
\begin{eqnarray}
	\frac{d^2R_{\omega l}}{dr_{*}^2}+\left(\omega^2-V_{\text{eff}}\right)R_{\omega l}=0.\label{radial equation}
\end{eqnarray}
Here $r_*$ is the proper radial distance given in (S2) of Section \ref{section2} and the effective potential is given by
\begin{eqnarray}
V_{\text{eff}}(r)=m_0^2+\frac{l(l+1)}{r^2}+\frac{b_0}{2r^3}
\end{eqnarray}

For convenience, we introduce two sets of basis functions $\psi^{L}_{\omega lm}$ and $\psi^{R}_{\omega lm}$ which are given by
\begin{eqnarray}
\psi^{L/R}_{\omega lm}(t,r,\theta,\varphi)=\frac{1}{\sqrt{4\pi|\omega|}}e^{-i\omega t}Y_{l m}(\theta,\varphi)\frac{R^{L/R}_{\omega l}(r)}{r}, \label{mode funtion}
\end{eqnarray}
where
\begin{eqnarray}
R_{\omega l}^{L}=\left\{\begin{array}{lr}
e^{i\sqrt{\omega^2-m_0^2}r_{*}}+\mathcal{R}^{L}_{\omega l}e^{-i\sqrt{\omega^2-m_0^2} r_{*}}, & r_{*}\rightarrow-\infty,\\\
\mathcal{T}^{L}_{\omega l}e^{i\sqrt{\omega^2-m_0^2} r_{*}},     & r_{*}\rightarrow\infty,\\
\end{array}\right.\quad
R_{\omega l}^{R}=\left\{\begin{array}{lr}
e^{-i\sqrt{\omega^2-m_0^2} r_{*}}+\mathcal{R}^{R}_{\omega l}e^{i\sqrt{\omega^2-m_0^2} r_{*}}, & r_{*}\rightarrow\infty,\\\
\mathcal{T}^{R}_{\omega l}e^{-i\sqrt{\omega^2-m_0^2} r_{*}},     & r_{*}\rightarrow-\infty.\\
\end{array}\right.\label{boundary condition1}
\end{eqnarray}
Here, $R^{L}_{\omega l}$ and $R^{R}_{\omega l}$ are two sets of basis solutions of radial equation (\ref{radial equation}), while $\mathcal{R}^{L/R}_{\omega l}$ and $\mathcal{T}^{L/R}_{\omega l}$ represent the corresponding reflection and transmission amplitudes.

According to \cite{Lanir:2017oia,Balakumar:2022yvx}, the symmetrized two-point correlation function in a thermal state at temperature $T=\kappa/2\pi$ can be expressed as
\begin{eqnarray}
	G(x,x'):=\int_{0}^{\infty}dk\sum_{l=0}^{\infty}\sum_{m=-l}^{m=l}\coth\left(\frac{\pi\omega(k)}{\kappa}\right)\left(\{\psi^{L*}_{\omega(k) lm}(x),\psi^{L}_{\omega(k) lm}(x')\}+\{\psi^{R*}_{\omega(k) lm}(x),\psi^{R}_{\omega(k) lm}(x')\}\right), \label{thermal two point}
\end{eqnarray}
where $\omega(k)=\sqrt{k^2+m_0^2}$. Here the presence of a coth factor indicates that the system is in a state of thermal equilibrium and
\begin{eqnarray}
\{X(x),Y(x')\}=\frac{1}{2}\left(X(x)Y(x')+Y(x)X(x')\right).
\end{eqnarray}

To compute the expectation value of the energy-momentum tensor, we need to take the limit $x\rightarrow x'$ but this leads to a divergence in (\ref{thermal two point}). This is because the singular structure of the two-point correlation function is described by the Hadamard condition \cite{Kay:1988mu}. Therefore, we need to consider renormalization. Here we employ the point-splitting method, which involves first taking $x$ and $x'$ as two distinct points, subtracting the divergent term, and then taking the limit $x\rightarrow x'$. We separate the two points in the $t$-direction, and we have

\begin{eqnarray}
x=\left(t,r,\theta,\varphi\right),\quad x'=\left(t+\epsilon,r,\theta,\varphi\right).
\end{eqnarray}
Using the relation $\psi^{L/R}_{\omega lm}(x')=\psi^{L/R}_{\omega lm}(x)e^{-i\omega\epsilon}$, the renormalized two-point correlation function can be written as
\begin{eqnarray}
G_{\text{rem}}(x)=\lim_{\epsilon\rightarrow 0}\int_{0}^{\infty} F(x,k)\cos\left(\omega(k)\epsilon\right)dk-L(x,\epsilon),\label{Grem}
\end{eqnarray}
where 
\begin{eqnarray}
F(x,k)=\sum_{l=0}^{\infty}\sum_{m=-l}^{m=l}\coth\left(\frac{\pi\omega(k)}{\kappa}\right)\left(\{\psi^{L*}_{\omega(k) lm}(x),\psi^{L}_{\omega(k) lm}(x)\}+\{\psi^{R*}_{\omega(k) lm}(x),\psi^{R}_{\omega(k) lm}(x)\}\right),
\end{eqnarray}
and the counter term $L(x,\epsilon)$, which is determined by the Hadamard condition \cite{Decanini:2005eg}, takes the form
\begin{eqnarray}
L(x,\epsilon)=\frac{b(x)}{\epsilon^2}+\frac{c(x)}{\epsilon}+d(x)\log(|\sigma|)+e(x)+O(\epsilon),\label{L}
\end{eqnarray}
where 
\begin{eqnarray}
	b(r_0)=-\frac{1}{4\pi^2},\quad c(r_0)=0,\quad d(r_0)=\frac{m_0^2}{8\pi^2},\quad e(r_0)=-\frac{m_0^2\log 2}{16\pi^2}.
\end{eqnarray}
As $\epsilon\rightarrow0$, the first term on the right-hand side of (\ref{Grem}) leads to divergence through integration, while the divergence of the second term stems from the analytical expression (\ref{L}). In practice, since we need to compute the first term numerically, this form is not convenient to handle. Using the identities \cite{Anderson:1994hg,Levi:2016paz}
\begin{eqnarray}
\int_{0}^{\infty}\omega^3\cos(\omega\epsilon)d\omega&&=\frac{6}{\epsilon^4},\nonumber\\
\int_{0}^{\infty}\omega\cos(\omega\epsilon)d\omega&&=-\frac{1}{\epsilon^2},\nonumber\\
\int_{0}^{\infty}\log(\sigma)\cos(\omega\epsilon)d\omega&&=-\frac{\pi}{2\epsilon},\nonumber\\
\int_{0}^{\infty}\frac{1}{\omega+\mu e^{-\gamma}}\cos(\omega\epsilon)d\omega&&=-\log(\mu\epsilon)+O(\epsilon),
\end{eqnarray}
The renormalized two-point correlation function (\ref{Grem}) can be reformulated as
\begin{eqnarray}
G_{\text{rem}}(x)=\int_{0}^{\infty} \left[F(x,k)-\frac{k}{\sqrt{k^2+m_0^2}}F_{\text{sing}}(x,\omega(k))\right]dk
-\int_{0}^{m}F_{\text{sing}}(x,\omega(k))d\omega-e(x),
\end{eqnarray}
where
\begin{eqnarray}
F_{\text{sing}}(x,\omega)=-b(x)\omega-\frac{2}{\pi}c(x)\log(\omega)-d(x)\frac{1}{\omega+e^{-\gamma}}.
\end{eqnarray}
Thus, the analytical divergence in (\ref{L}) is transformed into an integral form. Similarly, we can obtain the expected value of $\phi_{;u}\phi_{;v}$, which can be expressed as
\begin{eqnarray}
G_{uv,\text{rem}}(x)=\int_{0}^{\infty} F_{uv,reg}(x,k)dk
-\int_{0}^{m_0}F_{uv,\text{sing}}(x,\omega(k))d\omega-e_{uv}(x), \label{GUV}
\end{eqnarray}
where
\begin{eqnarray}
	F_{uv,\text{reg}}(x,k)=F_{uv}(x,k)-\frac{k}{\sqrt{k^2+m_0^2}}F_{uv,\text{sing}}(x,\omega(k)) \label{freg}
\end{eqnarray}
and
\begin{eqnarray}
F_{uv}(x,k)=\sum_{l=0}^{\infty}\sum_{m=-l}^{m=l}\coth\left(\frac{\pi\omega(k)}{\kappa}\right)\left(\{\psi^{L*}_{\omega(k) lm,u}(x),\psi^{L}_{\omega(k) lm,v}(x)\}+\{\psi^{R*}_{\omega(k) lm,u}(x),\psi^{R}_{\omega(k) lm,v}(x)\}\right). \label{FUV}
\end{eqnarray}
Here
\begin{eqnarray}
F_{uv,\text{sing}}(x,\omega)=\frac{1}{6}a_{uv}(x)\omega^3-b_{uv}(x)\omega-\frac{2}{\pi}c_{uv}(x)\log(\omega)-d_{uv}(x)\frac{1}{\omega+e^{-\gamma}}, \label{fsinguv}
\end{eqnarray}
and
\begin{eqnarray}
	&&a_{tt}=\frac{3}{2\pi^2},\quad b_{tt}=\frac{m_0^2}{8\pi^2},\quad c_{tt}=0,\quad
	d_{tt}=\frac{1+20m_0^4r^6}{640\pi^2r^6},\quad e_{tt}=-\frac{(1+20m_0^4r^6)(-3+\log 2)}{1280\pi^2r^6},\nonumber\\
	&&a_{r_*r_*}=\frac{1}{2\pi^2},\quad b_{r_*r_*}=-\frac{1-3m_0^2r^3}{24\pi^2r^3},\quad
	c_{r_*r_*}=0,\quad d_{r_*r_*}=-\frac{1-40m_0^2r^3+60m_0^4r^6}{1920\pi^2r^6},\nonumber\\
	&&e_{r_*r_*}=\frac{-3+60m_0^4r^6(-1+\log 2)+\log 2-40m_0^2r^3\log 2}{3840\pi^2r^6},\nonumber\\
	&&a_{\varphi\varphi}=\frac{r^2}{2\pi^2},\quad b_{\varphi\varphi}=\frac{1+6m_0^2r^3}{48\pi^2r},\quad
	c_{\varphi\varphi}=0,\quad
	d_{\varphi\varphi}=-\frac{-1+10m_0^2r^3+30m_0^4r^6}{960\pi^2r^4},\quad\nonumber\\ &&e_{\varphi\varphi}=\frac{3-60m_0^4r^6(-1+\log 2)-20m_0^2r^3\log 2+\log 4}{3840\pi^2r^4}.
\end{eqnarray}
Based on this, the renormalized energy-momentum tensor can be expressed as
\begin{eqnarray}
	T_{uv,\text{rem}}=G_{uv,\text{rem}}-\frac{1}{2}g_{uv}g^{\rho\sigma}G_{\rho\sigma,\text{rem}}-\frac{1}{2}m^2_0g_{uv}G_{\text{rem}}+\frac{v_1}{4\pi^2}g_{uv}. \label{remTuv}
\end{eqnarray}
Here the last term on the right-hand side corresponds to the trace anomaly \cite{Wald:1978pj}.

\section{The Renormalized Stress-Energy Tensor and Morris-Thorne Conditions}\label{section4}

In this section, we will numerically compute the renormalized energy-momentum tensor at the throat of a zero-tidal wormhole for a massive scalar field in a thermal state with temperature $T$. By varying the dimensionless scalar field mass $m_0b_0$ and the dimensionless temperature $b_0T$, we will investigate whether the quantum energy-momentum tensor satisfies the Morris-Thorne condition.

Here, we proceed to compute the renormalized energy-momentum tensor, whose explicit form is given by (\ref{remTuv}). For this purpose, we will employ the pragmatic mode-sum regularization method \cite{Levi:2015eea,Levi:2016paz}. We will present the detailed computational procedure for $G_{r_*r_*}$, while the remaining terms can be obtained through an entirely analogous approach.

The expression for $G_{r_*r_*}$ is given in (\ref{GUV}), we only need to compute $F_{r_*r_*}$ numerically, as provided in (\ref{FUV}). Using (\ref{mode decompose}), we have
\begin{eqnarray}
F_{r_{*}r_{*}}(x,k)=\sum_{l=0}^{\infty}\sum_{m=-l}^{m=l}\frac{1}{4\pi\omega(k)}\vert Y_{lm}(\theta,\varphi)\vert^2\coth\left(\frac{\pi\omega(k)}{\kappa}\right)\left(\left|\frac{d}{dr_*}\left(\frac{R^{L}_{\omega(k) l}}{r}\right)\right|^2+\left|\frac{d}{dr_*}\left(\frac{R^{L}_{\omega(k) l}}{r}\right)\right|^2\right).\label{Frstarstar}\nonumber\\
\end{eqnarray}
To evaluate the summation over $m$ in (\ref{Frstarstar}), we employ the following identity
\begin{eqnarray}
\sum_{m=-l}^{m=l}\left|Y_{lm}(\theta,\varphi)\right|^2=\frac{2l+1}{4\pi}.
\end{eqnarray}
‌To obtain $R^{L}_{\omega(k) l}$ and $R^{R}_{\omega(k) l}$, we numerically solve (\ref{radial equation}) subject to boundary condition (\ref{boundary condition1}) for $k$ ranging from 1/1000 to 60‌ in steps of ‌1/100‌. For each fixed $k$‌, we sum over $l$ in (\ref{Frstarstar}) until its contribution is less than $10^{-12}$‌. 
\begin{figure}[h]
	\begin{center}
		\includegraphics[scale=0.4]{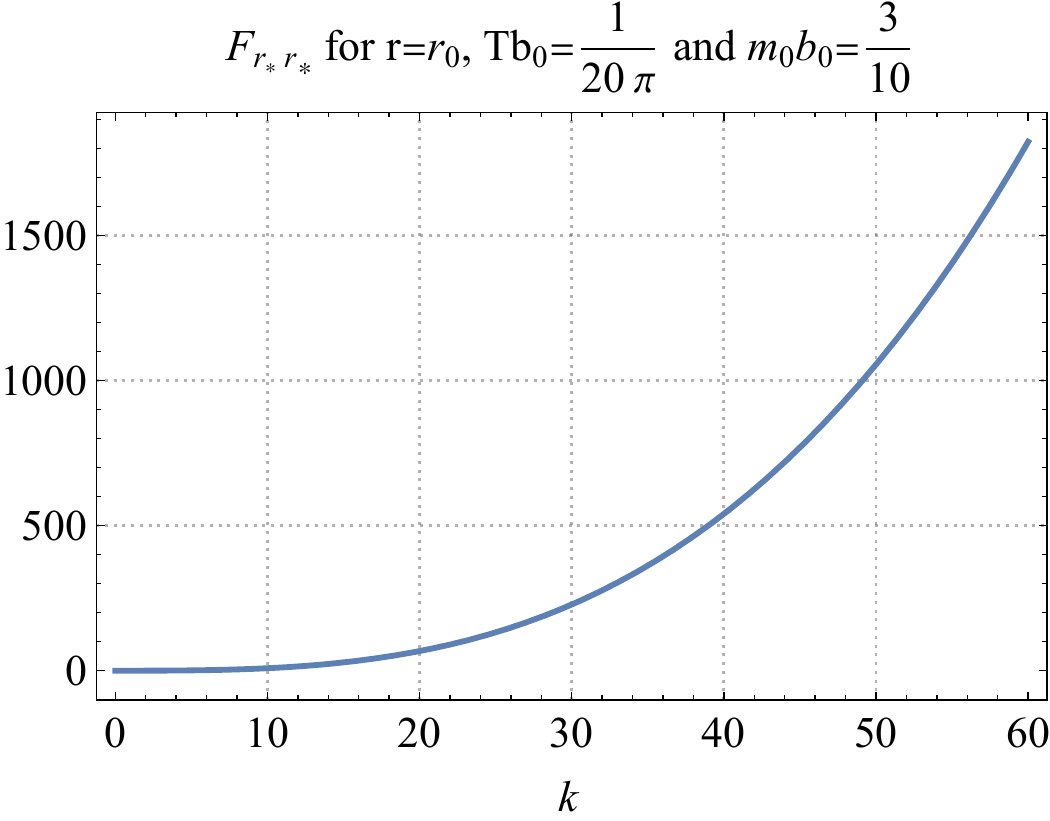}
	\end{center}
	\caption{The curve in this figure displays the value of $F_{r_{*}r_{*}}$ for $r=r_0$, $Tb_0=1/20\pi$ and $m_0b_0=3/10$ in the zero-tidal wormhole.}
	\label{figure1}
\end{figure}

Fig.\ref{figure1} displays $F_{r_{*}r_{*}}$ for $r=r_0$, $Tb_0=1/20\pi$ and $m_0b_0=3/10$. From Fig.\ref{figure1}‌, we observe that $F_{r_{*}r_{*}}$ increases as $k$ increases, and thus directly integrating with respect to $k$ would lead to a divergent result. This is because the two-point correlation function satisfies the Hadamard condition‌\cite{Kay:1988mu}, which leads to divergence, and the divergence can be described by (\ref{fsinguv}). The observation in Fig.\ref{figure1} that $F_{r_{*}r_{*}}$ increases with $k$ confirms the presence of such a divergence.

\begin{figure}[h]
	\begin{center}
		\includegraphics[scale=0.4]{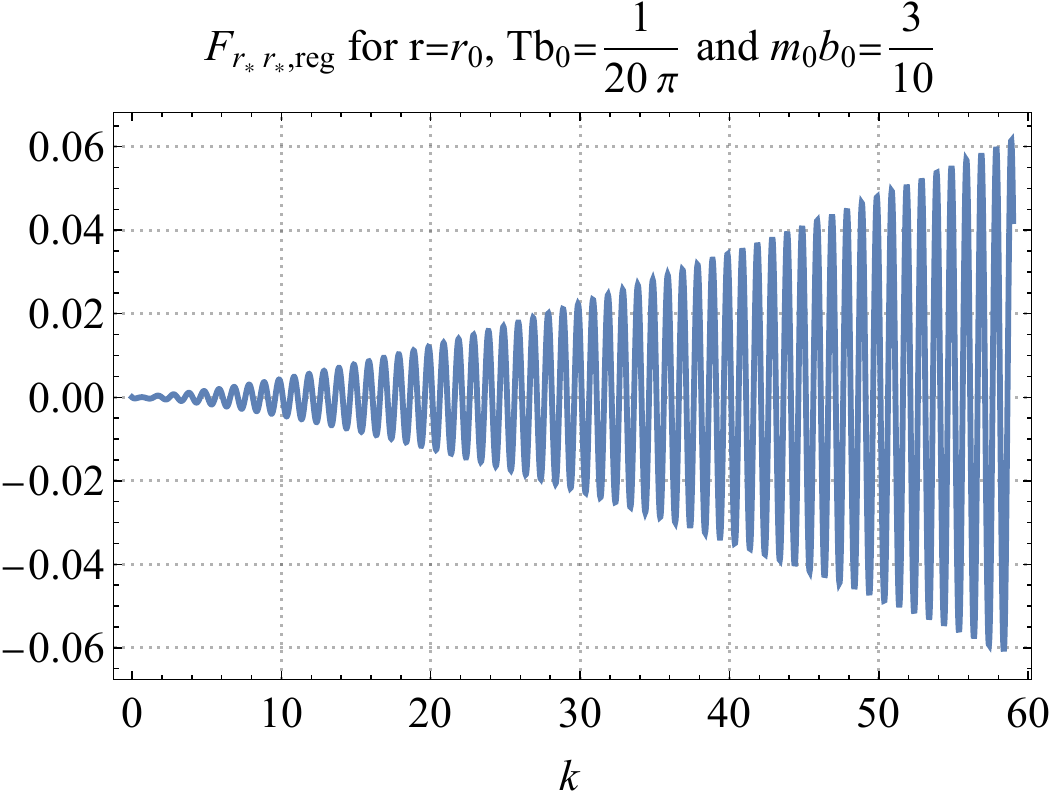}
	\end{center}
	\caption{The curve in this figure displays the value of $F_{r_{*}r_{*},\text{reg}}$ for $r=r_0$, $Tb_0=1/20\pi$ and $m_0b_0=3/10$ in the zero-tidal wormhole.}
	\label{figure2}
\end{figure}
Fig.\ref{figure2} displays $F_{r_{*}r_{*},\text{reg}}$ for $r=r_0$, $Tb_0=1/20\pi$ and $m_0b_0=3/10$, where $F_{r_{*}r_{*},\text{reg}}$ is given in (\ref{freg}). $F_{r_{*}r_{*},\text{reg}}$ corresponds to the result after subtracting the divergent term from  $F_{r_{*}r_{*}}$. Comparing Fig.\ref{figure1} and Fig.\ref{figure2}‌, we observe that the value of $F_{r_{*}r_{*}}$ at $k=60$‌ decreases from $1900$ in Fig.\ref{figure1} to $0.06$ for $F_{r_{*}r_{*},\text{reg}}$ at the same $k$ in Fig.\ref{figure2}, indicating that we have successfully removed the divergent term. After removing the divergent term, one might assume that the integral of the first term on the right-hand side of (\ref{GUV}) has already converged. However, this is not the case, and we denote this integral as
\begin{eqnarray}
	H_{r_*r_*,\text{reg}}(x,k)=\int_{0}^{k} F_{r_*r_*,reg}(x,k_1)dk_1
\end{eqnarray}

\begin{figure}[h]
	\begin{center}
		\includegraphics[scale=0.4]{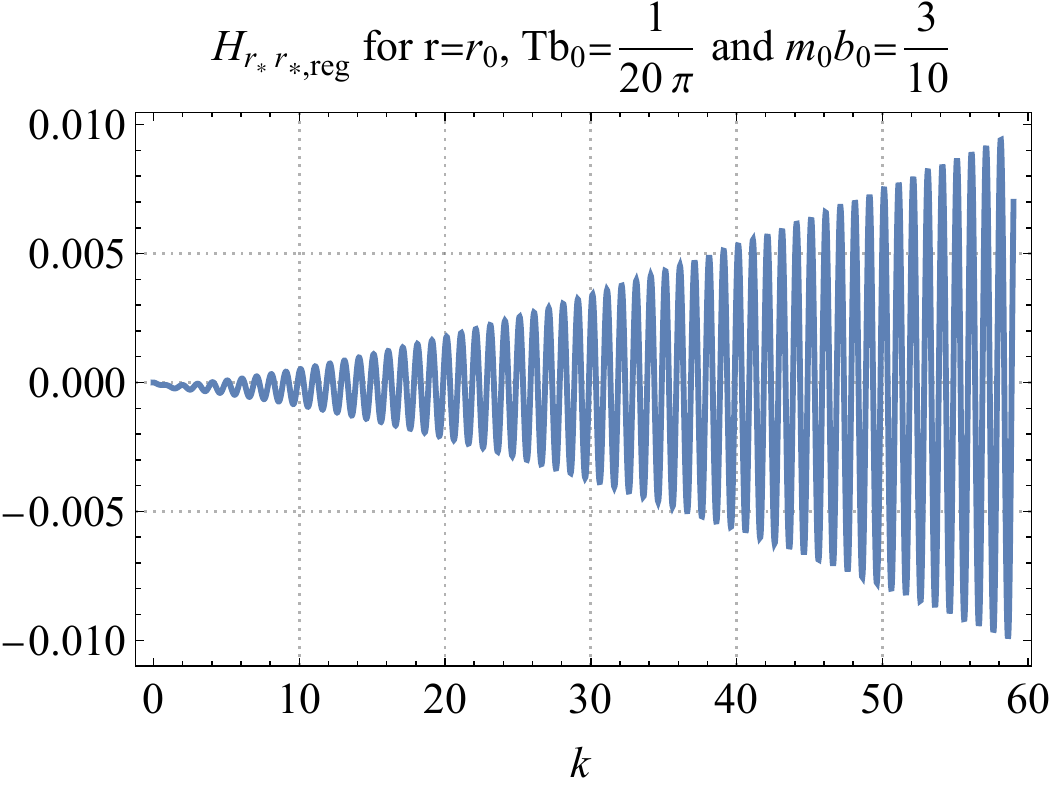}
	\end{center}
	\caption{The curve in this figure displays the value of $H_{r_{*}r_{*},\text{reg}}$ for $r=r_0$, $Tb_0=1/20\pi$ and $m_0b_0=3/10$ in the zero-tidal wormhole.}
	\label{figure3}
\end{figure}

Fig.\ref{figure3} displays $H_{r_{*}r_{*},\text{reg}}$ for $r=r_0$, $Tb_0=1/20\pi$ and $m_0b_0=3/10$. From Fig.\ref{figure3}, we observe that the integral over $k$ does not converge, but instead exhibits strong oscillatory behavior.‌ The origin of these oscillations is pointed out in \cite{Levi:2015eea,Levi:2016paz}. Roughly speaking, this is due to the two-point correlation function satisfying the Hadamard condition \cite{Kay:1988mu}‌, which gives rise to non-local singularities, and these singularities give rise to the oscillatory behavior in $H_{r_{*}r_{*},\text{reg}}$. This leads to the fact that we are effectively dealing with an generalized integral.

To handle these oscillations, we employ the self-cancellation method developed in \cite{Levi:2015eea,Levi:2016paz}. For a given integrand $h(x)$, we define
\begin{eqnarray}
	H(x)=\int_{0}^{x}h(x_1)dx_1.
\end{eqnarray}
If $H(x)$ exhibits periodic oscillations with wavelength $\lambda$, we define
\begin{eqnarray}
	T_{\lambda}[H(x)]=\frac{H(x)+H(x+\frac{\lambda}{2})}{2}.
\end{eqnarray}
By using operation $T_{\lambda}$, we can eliminate the oscillation and obtain a convergent result. In practice, we encounter multiple oscillations, and each oscillation requires several operations to eliminate. After this series of operations, we obtain a convergent result as follows
\begin{eqnarray}
\lim_{x\rightarrow\infty}(T_{\lambda_1})^{n_1}(T_{\lambda_2})^{n_2}...(T_{\lambda_m})^{n_m}[H(x)].
\end{eqnarray}
We apply the self-cancellation method to address the oscillations in $H_{r_{*}r_{*},\text{reg}}$‌, and by further taking into account the second and third terms on the right-hand side of (\ref{GUV}), we define  
\begin{eqnarray}
	G_{r_{*}r_{*},\text{reg},k}(x,k):=(T_{\lambda_1})^{n_1}(T_{\lambda_2})^{n_2}...(T_{\lambda_m})^{n_m}[H_{r_{*}r_{*},\text{reg}}(x,k)]-\int_{0}^{m_0}F_{uv,\text{sing}}(x,\omega)d\omega-e_{uv}(x).
\end{eqnarray}

\begin{figure}[h]
	\begin{center}
		\includegraphics[scale=0.4]{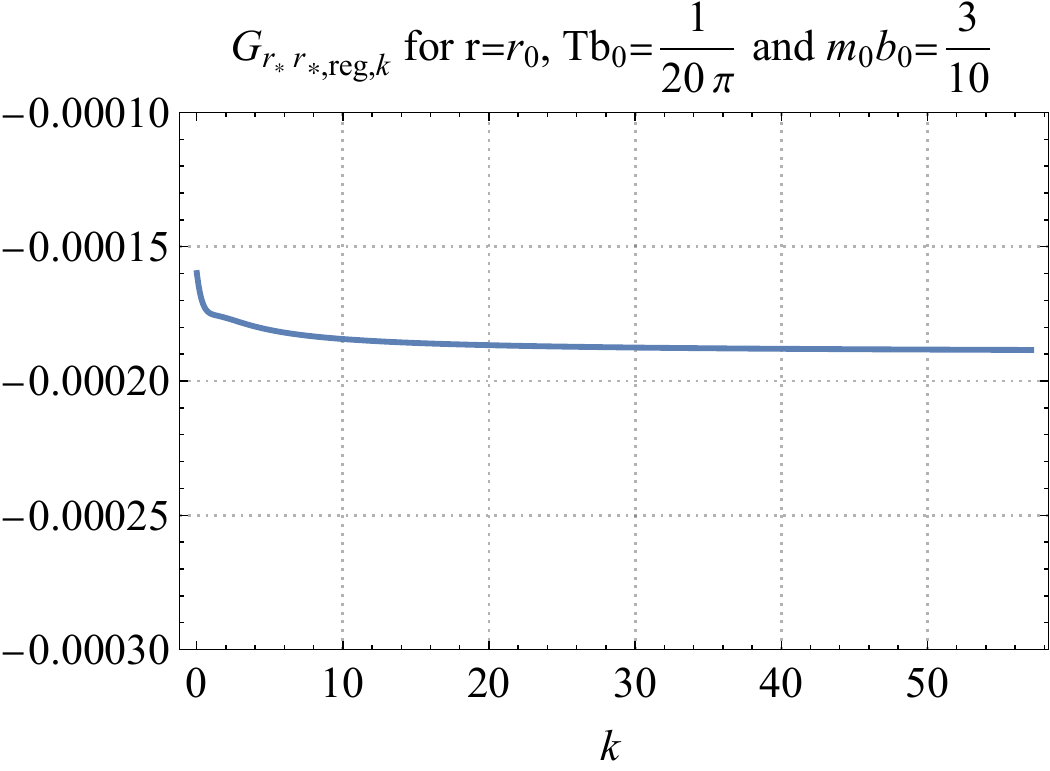}
	\end{center}
	\caption{The curve in this figure displays the value of $G_{r_{*}r_{*},\text{reg},k}$ for $r=r_0$, $Tb_0=1/20\pi$ and $m_0b_0=3/10$ in the zero-tidal wormhole.}
	\label{figure4}
\end{figure}

Fig.\ref{figure4} displays $G_{r_{*}r_{*},\text{reg},k}$ for $r=r_0$, $Tb_0=1/20\pi$ and $m_0b_0=3/10$.  From Fig.\ref{figure4}, we observe that that $G_{r_{*}r_{*},\text{reg},k}$ converges rapidly with increasing $k$, indicating that the self-cancellation method successfully eliminated the oscillations. Thus, the value of $G_{r_{*}r_{*},\text{reg}}$ is obtained, and other components can be calculated through similar methods.

Using this approach, we can compute all components of the quantum energy-momentum tensor $T_{uv,\text{rem}}$. To verify whether they satisfy the Morris-Thorne conditions, we define
\begin{eqnarray}
\tau_{\text{rem}}=-T_{r_*r_*,\text{rem}}(r_0),\label{QMT1}\\
\eta_{\text{rem}}=-T_{\text{tt},\text{rem}}(r_0)-T_{r_*r_*,\text{rem}}(r_0).\label{QMT2}
\end{eqnarray}

\begin{figure}[h]
	\begin{center}
		\includegraphics[scale=0.4]{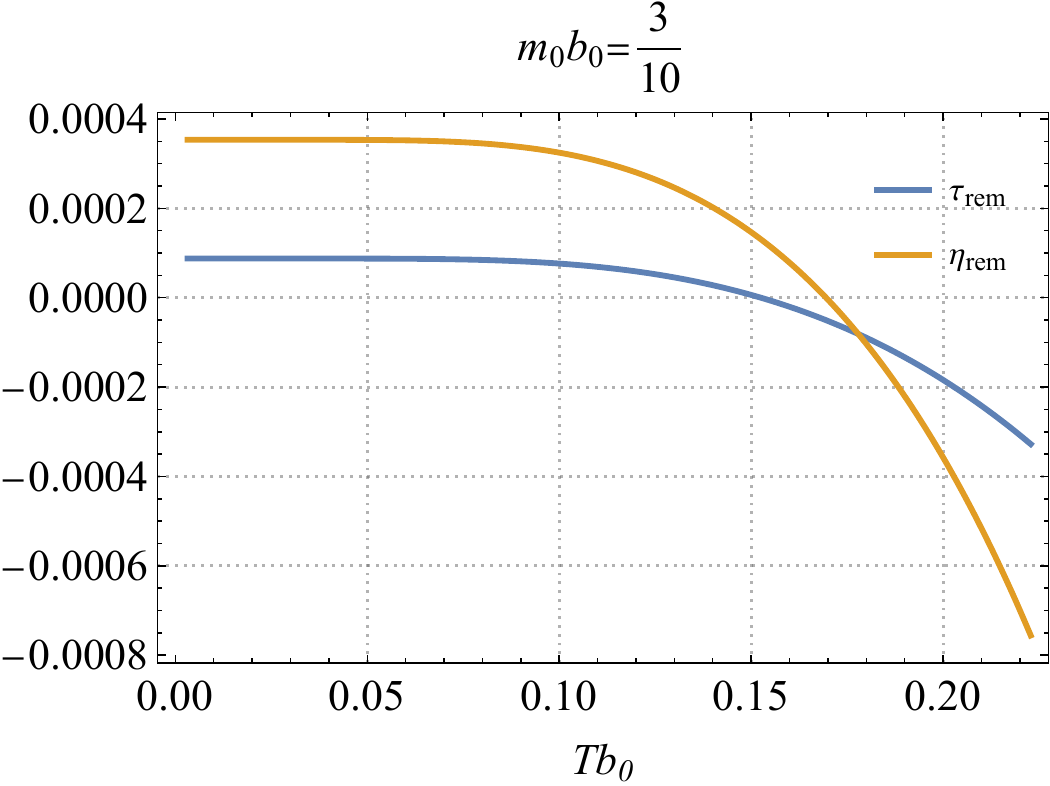}
	\end{center}
	\caption{The two curves in this figure display the value of $\tau_{\text{rem}}$ and $\eta_{\text{rem}}$ for $r=r_0$ and $m_0b_0=3/10$ in the zero-tidal wormhole.}
	\label{figure5}
\end{figure}
Fig.\ref{figure5} displays $\tau_{\text{rem}}$ and $\eta_{\text{rem}}$ for $r=r_0$ and $m_0b_0=3/10$. Fig.\ref{figure5} shows that at low temperatures $Tb_0$, both $\tau_{\text{rem}}$ and $\eta_{\text{rem}}$ remain positive, satisfying the Morris-Thorne conditions. As  $Tb_0$ increases, however, they decrease monotonically and eventually become negative, resulting in violation of the Morris-Thorne conditions.

\begin{figure}[h]
	\begin{center}
	\includegraphics[scale=0.4]{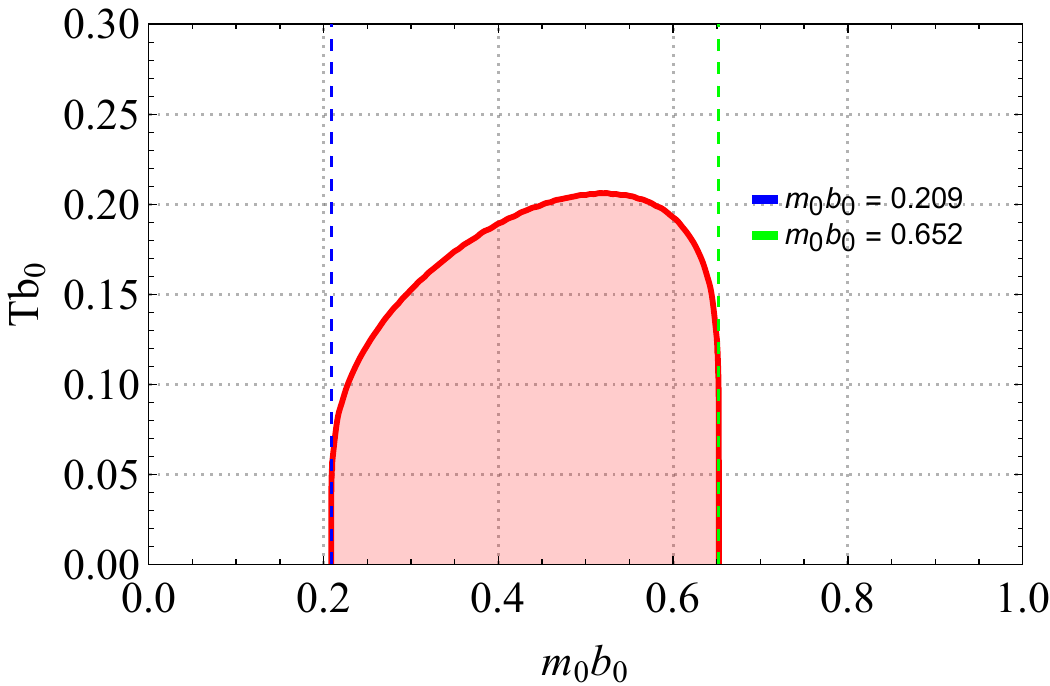}
	\end{center}
	\caption{This figure shows the dependence of the Morris-Thorne conditions on the dimensionless temperature $Tb_0$ and dimensionless scalar field mass $m_0b_0$, with red regions highlighting where these conditions are met.}
	\label{figure6}
\end{figure}

To investigate whether the quantum energy-momentum tensor of a massive scalar field in a thermal state can satisfy the Morris-Thorne conditions, we explore the two-dimensional parameter space of dimensionless temperature $Tb_0$ and dimensionless scalar field mass $m_0b_0$ to identify the regimes where quantum energy-momentum tensor can support wormhole spacetime structures. In Fig.\ref{figure6}, we examine the Morris-Thorne conditions within the parameter space spanned by dimensionless temperature $Tb_0$ and dimensionless scalar field mass $m_0b_0$. There exist two critical masses, $m_1b_0=0.209$ and $m_2b_0=0.652$. When $m_0b_0<m_1b_0$ or $m_0b_0>m_2b_0$, the quantum vacuum energy-momentum tensor of the thermal state fails to satisfy the Morris-Thorne condition at any temperature, thus failing to support the wormhole throat structure. For $m_0b_0\in(m_1b_0,m_2b_0)$, there exists a critical dimensionless temperature $Tb_0$, dependent on the dimensionless mass $m_0b_0$, below which the Morris-Thorne condition can be satisfied.

\section{Conclusions} \label{section5}

In this paper, we investigate the renormalized energy-momentum tensor at the throat of a zero-tidal wormhole for a massive scalar field in a thermal state. Employing the Hadamard renormalization framework, we perform numerical calculations of the renormalized stress-energy tensor using the pragmatic mode-sum regularization method developed by Levi and Ori \cite{Levi:2015eea,Levi:2016paz}. By varying the temperature of the thermal state and the scalar field mass, we investigate whether the quantum energy-momentum tensor can satisfy the Morris-Thorne conditions. In Fig.\ref{figure6}, the parameter region where Morris-Thorne conditions are satisfied is shown in the $m_0b_0-Tb_0$ parameter space.

Fig.\ref{figure6} reveals the existence of two critical masses $m_1b_0=0.209$ and $m_2b_0=0.652$. When the dimensionless scalar field mass $m_0b_0$ falls outside the interval $(m_1b_0,m_2b_0)$, the Morris-Thorne conditions cannot be satisfied regardless of temperature variations. For $m_0b_0\in(m_1b_0,m_2b_0)$, we observe a critical temperature $T_{cr}b_0$, which depends on $m_0b_0$, the Morris-Thorne conditions can only be satisfied below this critical temperature. The reason for the existence of a critical temperature is that, as we observe from Fig.\ref{figure4}, the quantum energy-momentum tensor tends to violate the Morris-Thorne condition as the temperature increases.‌ 

Our results indicate that only within the compact parameter region shown in Fig.\ref{figure6} can a quantum massive scalar field satisfy the Morris-Thorne conditions, which implies that the energy-momentum tensor of the quantum scalar field tends to sustain the structure of the wormhole. Thus, one may construct a traversable wormhole by exploiting the quantum energy-momentum tensor within this parameter region. Our results also demonstrate that if the wormhole is supported by a thermal quantum state, its ambient temperature must have a critical upper bound.

\section{Acknowledgments}
This work is supported by the National Natural Science Foundation of China with Grant No. 12275087.

\end{document}